\documentclass{article}

\usepackage{arxiv}

\usepackage[utf8]{inputenc} 
\usepackage[english]{babel}
\usepackage[T1]{fontenc}    
\usepackage{hyperref}       
\usepackage{url}            
\usepackage{booktabs,array,multirow}       
\usepackage{amsfonts,amsmath,amssymb}       
\usepackage{cleveref}
\usepackage{nicefrac}       
\usepackage{microtype}      
\usepackage{graphicx}
\usepackage{natbib}
\usepackage{doi}
\usepackage{graphicx}
\usepackage[space]{grffile}
\usepackage{latexsym}
\usepackage{textcomp}
\usepackage{longtable}
\usepackage{tabulary}
\usepackage{etoolbox}
\usepackage[table,xcdraw]{xcolor}

\usepackage{siunitx}

\title{Combining data assimilation and machine learning to estimate parameters of a convective-scale model}

\author{{S. Legler} \\
	Meteorological Institute\\
	Ludwig-Maximilians-Universität\\
	Munich, Germany \\
	\texttt{S.Legler@physik.uni-muenchen.de} \\
	\And
	{T. Janji\'{c}} \\
	Meteorological Institute\\
	Ludwig-Maximilians-Universität\\
	Munich, Germany \\
}

\renewcommand{\shorttitle}{Legler and Janji\'{c}}

\hypersetup{
pdftitle={Combining data assimilation and machine learning to estimate parameters of a convective-scale model},
pdfauthor={S. Legler, T. Janji\'{c}},
pdfkeywords={convective scale data assimilation, parameter estimation, Bayesian neural network, EnKF, layer-wise relevance propagation},
}

\begin{document}
\maketitle
\selectlanguage{english}
\begin{abstract}
	Errors in the representation of clouds in convection permitting numerical weather prediction models can be introduced by different sources. These can be the forcing and boundary conditions, the representation of orography, the accuracy of the numerical schemes determining the evolution of humidity and temperature, but large contributions are due to the parametrization of microphysics and the parametrization of processes in the surface and boundary layers. These schemes typically contain several tunable parameters that are either not physical or only crudely known, leading to model errors. Traditionally, the numerical values of these model parameters are chosen by manual model tuning. More objectively, they can be estimated from observations by the augmented state approach during the data assimilation.
\\
Alternatively, in this work, we look at the problem of parameter estimation through an artificial intelligence lens by training two types of artificial neural networks (ANNs) to estimate several parameters of the one-dimensional modified shallow-water model as a function of the observations or analysis of the atmospheric state. Through perfect model experiments we show that Bayesian neural networks (BNNs) and Bayesian approximations of point estimate neural networks (NNs) are able to estimate model parameters and their relevant statistics. The estimation of parameters combined with data assimilation for the state decreases the initial state errors even when assimilating sparse and noisy observations. The sensitivity to the number of ensemble members, observation coverage and neural network size is shown. Additionally, we use the method of layer-wise relevance propagation to gain insight into how the ANNs are learning and discover that they naturally select only a few grid points that are subject to strong winds and rain to make their predictions of chosen parameters.
\end{abstract}


\section{Introduction}
In recent years machine learning (ML) has become a subject of interest in various research fields within atmospheric physics. Attempts of including ML into climate and weather modeling reach from using it to represent sub-grid processes in global climate models \citep{gorman18,rasp18,yuval20}, over replacing data assimilation (DA) by an artificial neural network (ANN) to emulate the ensemble Kalman filter \citep[EnKF,][]{cintra14}, to utilizing an ANN as a surrogate for the complete physical model \citep{brajard20} or for the model error \citep{farchi21} during the DA. 
 \citet{bonavita20} use ANNs to estimate model error tendencies in the Integrated Forecasting System (IFS) of the European Centre for Medium-Range Weather Forecasts (ECMWF) and show that they are able to emulate the main outcomes acquired by the weak-constraint four dimensional variational (4D-Var) algorithm. Furthermore, computational cost can be improved when including ML in the DA. For example, \citet{ruckstuhl21} used a convolutional neural network (CNN) to show that a hybrid of a CNN and the EnKF is able to decrease the analysis/background error, equivalent to results obtained by the quadratic programming ensembles \citep[QPEns]{janjic14} but with a reduced computational cost compared to that of the QPEns. Finally, ML approaches can also be improved with DA methods by replacing the backpropagation during the training with an adaptive EnKF \citep{trautner20}. 
 
 While there are many examples of using ML to enhance the analysis/forecast of the model state, deep learning for model parameter estimation is not well developed, especially on the convective scale. In \citet{yadav20} the coupling parameter of the two-level Lorenz-96 model \citep{lorenz05} is estimated as a function of the resolved, large-scale state variable using a Gaussian Process (GP) \citep{rasmussen06}. The GP was compared to two types of ANNs and a simple linear regression and outperformed the other methods in most of the experiments. Similarly, data assimilation has been successfully used in geosciences for estimation of the state from sparse and noise observations. However, when  parameters are jointly estimated with the state several problems arise, for example, parameters are not directly observed and therefore updated through cross-correlations which might not be accurate; parameter values often need to be within certain bounds therefore Gaussian assumptions of data assimilation algorithms are not valid and finally 
 to use data assimilation for parameter estimation, stochastic model
for the parameters needs to be pre-specified to keep the spread in parameters \citep{ruckstuhl18,RuckstuhlJanjic20}. In this study, we investigate a possibility of using data assimilation for the state estimation while using ML for parameter estimation in order to overcome some of the problems of augmented state approach for estimating parameters from observations via data assimilation.
 
Although ML algorithms show promising results in idealized test cases, they come with two major drawbacks. On one hand, ANNs typically do not provide an uncertainty with their predictions, which makes it hard to ascribe a confidence when using them in operational settings. On the other hand, they are still seen as black boxes that do not provide any insight into the functions they are trying to approximate. To tackle the latter problem \citet{toms20} introduced layer-wise relevance propagation (LRP) to the geosciences, which can be used to visualize how the ANN makes its prediction. \citet{labe21} utilized this method to disentangle relative influences on regional surface temperatures of aerosols and greenhouse gases in the atmosphere. The former drawback could be approached by using stochastic ANNs instead of their widely used deterministic counterpart.
The goal of this study is threefold. First, to estimate parameters of the convective-scale modified shallow water model from sparse and noisy observations using ML and DA. Second, to compare the predictions and statistics of stochastically trained Bayesian neural networks (BNNs) with ensemble of deterministically trained point estimate neural networks. And third, to visualize the decision making of the ANNs by applying LRP.

The paper is organized as follows.  The model and the DA for the state are described in Section~\ref{sec:model}. Section~\ref{sec:ML}  introduces two ML algorithms for parameter estimation, and their use in combination with DA. This is followed by an investigation of the performance of these two hybrid algorithms in state and parameter space in Section~\ref{sec:results}. A final discussion and some perspectives are presented in Section~\ref{sec:con}.

\section{Dynamical model and state estimation}\label{sec:model}
\subsection{Modified shallow water model}
For this study the same dynamical model, model parameters, and parameter bounds (\autoref{tab:bounds}) as in \citet{ruckstuhl18} were used to conduct the experiments. However, instead of using data assimilation with augmented state approach for parameter estimation, we estimate the parameters with ANNs. In the \emph{twin experiments} presented in this study the true state (\emph{nature run}) of the atmosphere is generated by the modified shallow water model \citep{wuersch14}. Synthetic observations are produced by adding random perturbations to the true state. This model is computationally inexpensive but still represents the key space and time scales of storm developments. It is based on the shallow water equations for the fluid velocity \( u \) and the fluid height \( h \) with a modification of the geopotential \( \phi \) to include conditional instability. Additionally, a variable for the rain \( r \) was added to mimic nature. The equations are as follows:
\begin{equation}
 \frac{\partial u}{\partial t} + u \frac{\partial u}{\partial x} + \frac{\partial (\phi + c^2r)}{\partial x} = \beta_u + D_u \frac{\partial^2 u}{\partial x^2}
 \label{eq:u}
\end{equation}
\begin{equation}
\phi =  \begin{cases}
      \phi_c & \text{if $h > h_c$}\\
      gh & \text{else}
    \end{cases}
\label{eq:phi}
\end{equation}
\begin{equation} 
\frac{\partial r}{\partial t} + u \frac{\partial r}{\partial x} =  D_r \frac{\partial^2 r}{\partial x^2} - \alpha r - \begin{cases}
      																			\delta \frac{\partial u}{\partial x} & \text{if $h>h_r$ and $\frac{\partial u}{\partial x}<0$}\\
																		             0 & \text{else}
																		    \end{cases}
\label{eq:r}
\end{equation}
\begin{equation}
\frac{\partial h}{\partial t} + \frac{\partial (uh)}{\partial x} = D_h \frac{\partial^2 h}{\partial x^2}
\label{eq:h}
\end{equation}
\begin{itemize}
	\item \( D_u, D_r, D_h : \) diffusion constants
	\item \( c^2 = g \times h_0 : \) gravity-wavespeed for absolute fluid layer \( h_0 ( h_0 < h_c ) \)
	\item \( \delta : \) production rate of rain
	\item \( \alpha : \) removal rate of rain
\end{itemize}
Convection is triggered by adding a low amplitude noise source \( \beta_u \) at random locations to the velocity at every model time step. When the fluid height \( h \) exceeds the threshold \(h_c\), which represents the level of free convection, the geopotential is replaced by a lower constant value \( \phi_c \). The gradient of the geopotential forces fluid to the regions of lower geopotential, which then builds up the fluid height in those regions. Once \( h \) reaches the threshold \( h_r \) rain is being produced by adding rainwater mass to the geopotential. The removal of rain is mimicked by a linear relaxation towards zero. 
For the experimental set-up a one dimensional grid of length 125 km with 250 grid points was used, which yields a state vector of the form:
\begin{equation}
\mathbf{x} =\begin{bmatrix}\mathbf{u}\\\mathbf{h}\\\mathbf{r}\end{bmatrix} \in \mathbb{R}^{750}.
\end{equation}
The model parameters which were chosen to be estimated are the rain removal rate \( \alpha \), the low constant value for the geopotential \( \phi_c \) and the threshold for the fluid height \( h_r \) while the other parameters were known during the experiments. Assuming all of the model parameters are constant in space the resulting parameter vector is:
\begin{equation}
\mathbf{\theta} =\begin{bmatrix}\alpha\\\phi_c\\h_r\end{bmatrix} \in \mathbb{R}^{3}.
\end{equation}
Observations are generated from the nature run every 60 model time steps by adding a Gaussian error to \( u \) and \( h \) and a lognormal error to \(r\) variable to keep its positivity. To simulate radar data only the grid points where \( r > 0.005 \) were observed. Furthermore, wind observations of \( 25 \% \) of the remaining grid points were added. The model parameters for the nature run are taken from uniform distributions. The upper and lower bounds of the uniform distributions for the model parameters as well as biases and standard deviations of the observational errors are summarized in \autoref{tab:bounds} and \autoref{tab:error} respectively.
\begin{table}[h!]
\centering
\scalebox{0.8}{
\begin{tabular}{|l|l|l|}
\hline
\rowcolor[HTML]{EFEFEF} 
Parameter & Lower Bound & Upper Bound \\ \hline
\(\alpha \)     & 0.0003      & 0.001       \\ \hline
\( \phi_c \)       & 899.7       & 899.9       \\ \hline
\( h_r \)         & 90.15       & 90.25       \\ \hline
\end{tabular}}
\caption{\label{tab:bounds}Lower and upper bounds for the uniform distributions of the model parameters}
\end{table}
\begin{table}[h!]
\centering
\scalebox{0.8}{
\begin{tabular}{|l|l|l|}
\hline
\rowcolor[HTML]{EFEFEF} 
Variable & Mean     & Standard Deviation \\ \hline
\(u\)        & 0       & 0.001       \\ \hline
\(h\)       & 0        & 0.02         \\ \hline
\(r\)        & 0.001 & 1e-7        \\ \hline
\end{tabular}}
\caption{\label{tab:error}Means and standard deviations for the  distributions of the observational errors}
\end{table}

\subsection{Stochastic ensemble Kalman filter}
Since the focus of this work is testing new algorithms for parameter estimation, a simple stochastic EnKF \citep{evensen94,evensen03} will be utilized for all experiments. It is based on the following cost function for each of the $N_{ens}$ ensemble members:
\begin{equation}
J(\mathbf{x}^{a,i}_t) = (\mathbf{x}^{f,i}_t-\mathbf{x}^{a,i}_t)^T\mathbf{P}^{-1}_t(\mathbf{x}^{f,i}_t-\mathbf{x}^{a,i}_t) + (\mathbf{y}^{i}_t-\mathbf{H}_t\mathbf{x}^{a,i}_t)^T\mathbf{R}^{-1}_t(\mathbf{y}^{i}_t-\mathbf{H}_t\mathbf{x}^{a,i}_t)
\label{eq:cost}
\end{equation}
where \(i, i=1...N_{ens}\) denotes one ensemble member, \( \mathbf{x}^{f/a,i}_t \) are the background and analysis states respectively, \(\mathbf{R}_t \) is the observation-error covariance matrix and \(\mathbf{H}_t\) denotes the observation operator, that maps the model states to the observation space. In this work we assume \(\mathbf{H}_t\) to be linear. \(\{\mathbf{y}^i_t\}\) represents an ensemble of observations acquired by perturbing the observation vector \( \mathbf{y}_t \) such that \(\mathbf{y}^i_t = \mathbf{y}_t + \mathbf{\epsilon}^i \). \( \mathbf{\epsilon}^i \) is a  perturbation taken from a distribution with a bias and a standard deviation that represent the observation error. The subscript \(t\) refers to the time when a DA cycle is being carried out, which usually corresponds to the time appropriate observations are available. For the rest of this chapter, the subscript \(t\) will be withheld. The forecast-error covariance matrix is generated with the ensemble of background states:
\begin{equation}
\mathbf{P} = \overline{(\mathbf{x}^{f,i}-\overline{\mathbf{x}^{f}})(\mathbf{x}^{f,i}-\overline{\mathbf{x}}^{f})^T}
\label{eq:P}
\end{equation}
where the overline denotes the average over the ensemble members. Minimizing \(J\) for each ensemble member yields the analysis ensemble:
\begin{equation}
\mathbf{x}^{a,i} = \mathbf{x}^{f,i} + \mathbf{P}\mathbf{H}^T(\mathbf{H}\mathbf{P}\mathbf{H}^T + \mathbf{R})^{-1}(\mathbf{y}^i - \mathbf{H}\mathbf{x}^{f,i})
\label{eq:analysis}
\end{equation}
with the Kalman gain \(\mathbf{K}= \mathbf{P}\mathbf{H}^T(\mathbf{H}\mathbf{P}\mathbf{H}^T + \mathbf{R})^{-1} \). In all experiments exhibited in this study \Cref{eq:analysis} was used to estimate only the atmospheric state. Once analysis ensemble and corresponding parameters (see Section \ref{sec:ML}) are estimated, nonlinear model  (\ref{eq:u}) - (\ref{eq:h}) would be used to obtain forecast ensemble \( \mathbf{x}^{f,i} \) for time \(t+1\).

\section{ML for parameter estimation} \label{sec:ML}
 The scientific objective of this study is to estimate three model parameters of the modified shallow water model as a function of the atmospheric state consisting of the atmospheric variables \(u,h,r\) using different types of ANNs.

\subsection{Types and architecture of ANNs}
Two types of ANNs are utilized - a \emph{deep ensemble of point estimate neural networks (NN)} and a \emph{Bayesian neural network (BNN)}. Both have an input size of 750 - three atmospheric variables for each of the 250 grid points - and an output size of three which corresponds to the three global, unknown parameters that are wished to be estimated. The term point estimate neural network is used in this work to refer to the standard type of neural network which given an input predicts one deterministic output. To quantify the uncertainty of the estimation produced by the NN the method of deep ensembles from \citet{lakshminarayanan17} is adopted. This is an easy implementable approach, where an ensemble of neural networks \( \{NN_k\}^{n_{NN}}_{k=1} \) consisting of \(n_{NN}\) members with the same architecture but random initial weights is trained independently. The definition of BNNs is not completely consistent across literature. We adopt its definition from \citet{jospin20} as a type of ANN "...built by introducing stochastic components into the network..." and trained using \emph{Bayesian inference} \citep{mackay92}. Stochastic components can either be introduced as probability distributions over the activation functions or over the weights, although for this study the latter one is utilized as this is the more common one. For a more detailed account about the differences between NNs and BNNs and how to utilize BNNs we refer to \citet{jospin20}. For both ANNs (\Cref{tab:architecture_nn,tab:architecture_bnn}) fully connected linear layers were chosen with additional batch normalization layers to accelerate the training \citep{ioffe15}. For the NN a dropout layer was added to reduce overfitting \citep{labach19} which was not necessary for the BNN. \emph{ReLU} was used as the activation function for all hidden layers of the NN. This was not possible for the BNN as it resulted in the  \emph{dying ReLU problem} \citep{lu20} which is a widely known phenomenon where ReLU neurons output 0 for all inputs. To combat this the \emph{LeakyReLU} was utilized for all hidden layers of the BNN. The activation functions are defined as:
\begin{equation}
ReLU(x) = max(0,x)
\end{equation}
\begin{equation}
LeakyReLU(x) = max(0,x) + 0.01 * min(0,x).
\end{equation}
The number of neurons for the hidden layers were optimized independently for the NN and the BNN and therefore differ from each other.
\begin{table}[]
\centering
\scalebox{0.7}{
\begin{tabular}{|c|c|c|}
\hline
\rowcolor[HTML]{EFEFEF} 
\multicolumn{3}{|c|}{\cellcolor[HTML]{EFEFEF}Architecture}          \\ \hline
\rowcolor[HTML]{EFEFEF} 
Type of Layer    & Size (input x output) & Activation Function          \\ \hline
\rowcolor[HTML]{FFFFFF} 
Linear           & 750 x 31              & ReLU                         \\ \hline
\rowcolor[HTML]{FFFFFF} 
Batch-Norm       & 31 x 31               & None                         \\ \hline
\rowcolor[HTML]{FFFFFF} 
Dropout (p=0.5)  & 31 x 31               & None                         \\ \hline
\rowcolor[HTML]{FFFFFF} 
Linear           & 31 x 19               & ReLU                         \\ \hline
\rowcolor[HTML]{FFFFFF} 
Linear           & 19 x 11               & ReLU                         \\ \hline
\rowcolor[HTML]{FFFFFF} 
Linear           & 11 x 3                & None \\ \hline
\rowcolor[HTML]{EFEFEF} 
\multicolumn{3}{|c|}{\cellcolor[HTML]{EFEFEF}Training}                  \\ \hline
\rowcolor[HTML]{FFFFFF} 
Optimizer        & \multicolumn{2}{c|}{\cellcolor[HTML]{FFFFFF}Adam}    \\ \hline
\rowcolor[HTML]{FFFFFF} 
Mini-Batch-Size  & \multicolumn{2}{c|}{\cellcolor[HTML]{FFFFFF}32}      \\ \hline
\rowcolor[HTML]{FFFFFF} 
Number of Epochs & \multicolumn{2}{c|}{\cellcolor[HTML]{FFFFFF}150}     \\ \hline
\end{tabular}}
\caption{\label{tab:architecture_nn}Architecture and training specifics of the \emph{NN}}
\end{table}
\begin{table}[h!]
\centering
\scalebox{0.7}{
\begin{tabular}{|c|c|c|c|c|c|}
\hline
\rowcolor[HTML]{EFEFEF} 
\multicolumn{3}{|c|}{\cellcolor[HTML]{EFEFEF}Architecture}       & \multicolumn{3}{c|}{\cellcolor[HTML]{EFEFEF}Stochastic Model}                                                \\ \hline
\rowcolor[HTML]{EFEFEF} 
Type of Layer & Size (input x output) & Activation Function          & \multicolumn{2}{c|}{\cellcolor[HTML]{FFFFFF}Priors \(p(W_i^{(k,l)})\)}         & \cellcolor[HTML]{FFFFFF}\( \mathcal{N}(0,1) \)  \\ \hline
\rowcolor[HTML]{FFFFFF} 
Batch-Norm    & 750 x 750             & None                         & \multicolumn{2}{c|}{\cellcolor[HTML]{FFFFFF}Variational distributions \(q_{\phi}(W_i^{(k,l)})\)} & \( \mathcal{N}(\mu,\sigma) \) \\ \hline
\rowcolor[HTML]{FFFFFF} 
Linear        & 750 x 20              & LeakyReLU                    & \multicolumn{3}{c|}{\cellcolor[HTML]{EFEFEF}Training}                                                        \\ \hline
\rowcolor[HTML]{FFFFFF} 
Linear        & 20 x 20               & LeakyReLU                    & Optimizer                                   & \multicolumn{2}{c|}{\cellcolor[HTML]{FFFFFF}Adam}              \\ \hline
\rowcolor[HTML]{FFFFFF} 
Linear        & 20 x 20               & LeakyReLU                    & Mini-Batch Size                             & \multicolumn{2}{c|}{\cellcolor[HTML]{FFFFFF}32}                \\ \hline
\rowcolor[HTML]{FFFFFF} 
Linear        & 20 x 3                & None & Number of Epochs                            & \multicolumn{2}{c|}{\cellcolor[HTML]{FFFFFF}3}                 \\ \hline
\end{tabular}}
\caption{\label{tab:architecture_bnn}Architecture, stochastic model and training specifics of the \(BNN\)}
\end{table}

\subsection{Data generation and training}
\label{chap:training}
To generate the input-output pairs for the training, validation, and test data-sets 100 000 sets of parameters are taken randomly from the uniform distributions specified in \autoref{tab:bounds}. 
For each set of parameters \Cref{eq:u,eq:phi,eq:r,eq:h} are solved for 1000 model time steps with a timestep discretization of \( \Delta t = 4 \). Snapshots in time at \(t = 1000\) of the state vectors are used as the input of the ANNs. The parameters used to generate those states are the corresponding outputs after rescaling them to [0,1]. From these 100 000 input-output pairs 90\% are used for training, 5\% for validation during the training and 5\% for testing. Additionally, the input samples are augmented during the training by adding perturbations taken from distributions with means and standard deviations corresponding to the observational error specified in \autoref{tab:error}. For each input sample three perturbed samples are added during the training resulting in a training size of \(360  000\). The NN is trained via stochastic gradient descent. Since computing the exact Bayesian posterior of the BNN is usually intractable, we utilize the optimization method of stochastic variational inference \citep{hoffman13} where a set of variational distributions is approximated to the exact Bayesian posteriors during the training. A widely used default for the prior weights of BNNs are normal distributions with mean 0 and standard deviation \( \sigma\) \citep{jospin20}. After evaluating the trained weights of the point estimate neural networks it seemed appropriate to set \( \sigma = 1\). To simplify training, the variational distributions are initialized as normal distributions as well such that during the training only the means and standard deviations have to be optimized. For both ANNs an adaptable learning rate \citep[Adam]{kingma14} was chosen with an initial value of 0.001. The NN is implemented and trained using only the library PyTorch \citep{paszke19} while for the BNN also the probabilistic programming language Pyro \citep{bingham18} which is built on top of PyTorch is used.

\subsection{Combining DA and ML}
\label{chap:da_ml}
All DA experiments presented in this study are conducted as twin experiments. The atmospheric state and model parameters of the nature run start from a sample taken from the test data-set. The state is propagated forward in time using the modified shallow water model while the parameters are kept constant. The background ensemble members start from different states. Whenever a DA cycle is performed according to \Cref{eq:analysis} the model  parameters are estimated according to one of the following set-ups.
\begin{enumerate}
	\item \textbf{true}: The true values of the parameters are known and used for the background state throughout all DA cycles.
	\item \textbf{random}: The true values of the parameters are not known and picked randomly from a the uniform distributions specified in \Cref{tab:bounds}.
	\item \textbf{NN}: The parameters are estimated using the observations (DA cycle = 0) or the analysis (DA cycle \(>\) 0) as input to an ensemble of NNs with 15 members trained according to \Cref{chap:training}.
    \item \textbf{BNN}\(_0\): The parameters are estimated using the observations (DA cycle = 0) or the analysis (DA cycle \(>\) 0) as input to a BNN (BNN\(_0\)) trained according to \Cref{chap:training}.
	\item \textbf{BNN}\(_0\)\textbf{+BNN}\(_t\): The parameters are estimated using the observations as input to BNN\(_0\) (DA cycle = 0) or to BNN\(_t\) (DA cycle \(>\) 0) trained according to \Cref{chap:remarks}.
\end{enumerate}
The parameter estimates are then used as parameters for the forward model simulations with the modified shallow water model for the next 60 model time steps until the next DA cycle is performed, and parameters are estimated again with ML algorithm.

\subsubsection{Remarks}
\label{chap:remarks}
\textbf{DA cycle = 0:} For the first DA cycle, the observations taken from the nature run are used as inputs for the ANNs. If a variable is only partially observed the unobserved grid points are interpolated with a quadratic interpolation for \(u\) and \(h\). For \(r\) the unobserved grid points are simply set to 0.\\
\textbf{NN:} At DA cycle \(>\) 0 each analysis ensemble member \(\mathbf{x}_t^{a,i} \) is used as input for each NN ensemble member resulting in \(15 \cdot N_{ens} \) parameter estimates. To obtain \(N_{ens}\) parameter vectors out of this ensemble a beta distribution is fitted to the ensemble of parameter estimates. The parameters for each state ensemble member is then taken from this beta distribution.\\
\textbf{BNN\(_t\):} This set-up is chosen to test the feasibility of online training during the DA cycle with a realistic number of forecast/analysis ensemble members. Since the training size of BNN\(_t\) is much smaller than that used for \(BNN_0\) it is necessary to reduce the number of trainable weights for \(BNN_t\). The first modification is to reduce the input size. Experiments from \citet{ruckstuhl18} indicate that the rain r and fluid height h are stronger correlated to the parameters than the wind u. Hence, instead of using all 3 atmospheric variables as the input, only r and h are used. Since BNN\(_t\) is trained from scratch at each DA cycle, the input size can be left variable. This allows to only train on those grid points that are actually observed. Additionally, because we assume the true parameters to be constant over the whole grid, it might not be necessary to use all observed grid points as input. Therefore if more than 62 gridpoints (about 25 \% of the whole grid) are observed, only those 62 grid points with the highest observed values for r are used. \(\tilde{\mathbf{x}}^{f,i}\) in \Cref{eq:BNNt_input} refers to this reduced background state. To further reduce the number of learnable weights, the number of neurons per hidden layer is decreased from 20 to 2. In total, this results in a maximum of around 540 learnable weights. The resulting input for the training is then given by
\begin{equation}
\label{eq:BNNt_input}
\mathbf{H}_t\tilde{\mathbf{x}}^{f,i}_{\tau}+\mathbf{\epsilon}^i
\end{equation}
with the observation operator \(\mathbf{H}_t\) and \(i=1,...,N_{ens}\). The 10 previous points in time \( \tau=t-9,...,t\) are used to increase the training size from \(N_{ens}\) to \(N_{ens} \cdot 10 \). The labels for these inputs are simply the model parameters \( \mathbf{theta}^i_{t-60}\) from the previous estimation. Noise corresponding to the observational error specified in \autoref{tab:error} was added during the training for the same reasons as for the training from \Cref{chap:training}. The stochastic model, optimizer, and mini-batch size are the same as for BNN\(_0\) but 9 training epochs were necessary to reach a minimum in the validation loss which is most likely caused by the reduced training size.

\section{Results} \label{sec:results}
\subsection{Diagnostics}
\label{chap:diagnostics}
Besides the standard diagnostic tools - Root Mean Squared Error (RMSE), ensemble spread (spread), and Coefficient of Determination (\(R^2\)) - we utilize \emph{LRP} in this study. LRP is a visualization tool, which takes a trained ANN and an ANN input sample as the input and produces a \emph{LRP heatmap} as the output. The LRP heatmap is a vector of the same size as the ANN input and those entries with higher numerical values can be interpreted as being more relevant for the ANN's prediction as the ones with lower values. This method has been introduced first to the field of computer vision by \citet{bach15} and the \emph{PyTorch} implementation used in this study is from \citet{bohle19}. For an in-depth explanation of the algorithm, we refer to \citet{toms20}, who recently introduced LRP to the geosciences. The underlying idea of LRP is to calculate a \emph{relevance} for each input pixel by taking a specific ANN output, which in this case would be either \(\alpha, \phi_c\) or \(h_r\), and propagating it back through the network according to a certain set of propagation rules. Applying LRP to the ANNs trained in this study could thus give insight into which grid points and atmospheric variables are most relevant for each of the three parameters.

\subsection{Evaluation}
\begin{figure}[h!]
\begin{center}
\includegraphics[width=1\columnwidth]{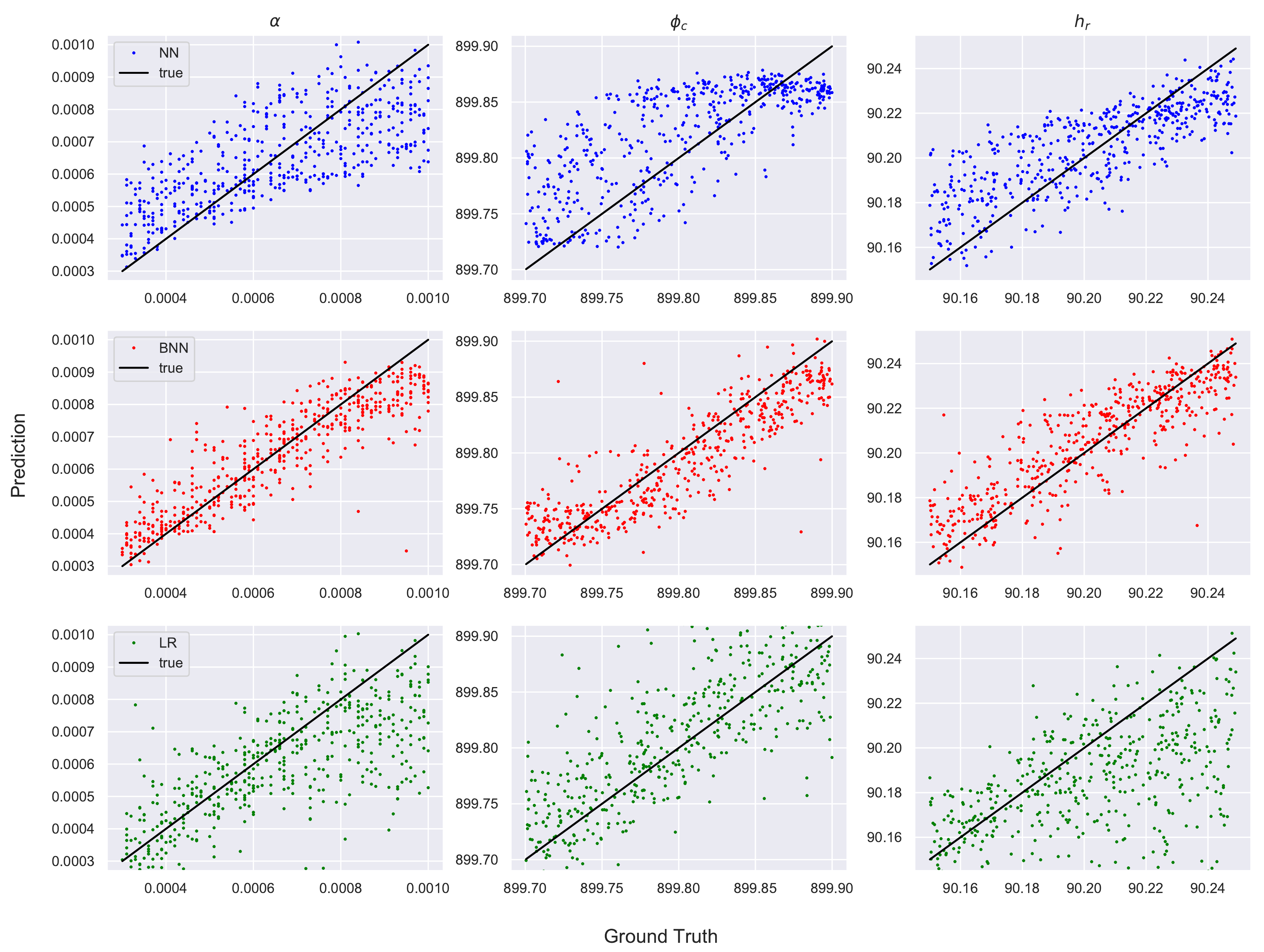}
\end{center}
\caption{Output of NN (blue dots), BNN (red dots), and LR (green dots) against corresponding ground truths and ideal output (black lines) of 500 samples}
    \label{fig:performance}
\end{figure}

For the first performance evaluation 500 samples from the test dataset  were used to estimate the parameters. For these experiments we assume a fully observed grid and no observation noise. Each output ensemble was averaged and then plotted against its corresponding ground truth (\Cref{fig:performance}). Additionally, as a baseline model a simple linear regression (LR) model was fitted to the same training data. As a benchmark, the ideal output was plotted as well, which corresponds to the black lines with slope 1. The BNN outperforms the NN as well as the LR in all three parameters while the LR has the lowest \(R^2\) scores for all parameters (\Cref{tab:r2}). While the BNN has similar \(R^2\) scores for the different parameters the NN's and LR's performance varies greatly between them. For \(h_r\) the LR performs even worse than a baseline model which would predict the average value of the parameter bounds for all inputs. The scatter plot in \Cref{fig:performance} emphasizes that both ANNs are slightly overestimating low parameter values while underestimating high ones while the LR predicts values that are greatly out of bounds for all parameters.
\begin{table}[h!]
\parbox[t]{.5\linewidth}{
\centering
\begin{tabular}{|l|l|l|l|}
\hline
\rowcolor[HTML]{EFEFEF} 
model & \(\alpha\) & \(\phi_c\)  & \(h_r\)     \\ \hline
\rowcolor[HTML]{FFFFFF} 
NN    & 0.53  & 0.44 & 0.62  \\ \hline
\rowcolor[HTML]{FFFFFF} 
BNN   & 0.79  & 0.74 & 0.75  \\ \hline
\rowcolor[HTML]{FFFFFF} 
LR    & 0.41  & 0.26 & -0.52 \\ \hline
\end{tabular}
\caption{\label{tab:r2}\(R^2\) of the parameter predictions plotted in \Cref{fig:performance} for NN, BNN and LR}
}
\parbox[t]{.5\linewidth}{
\centering
\begin{tabular}{|l|l|l|l|}
\hline
\rowcolor[HTML]{EFEFEF} 
model & \(\alpha\) & \(\phi_c\)  & \(h_r\)     \\ \hline
\rowcolor[HTML]{FFFFFF} 
NN    & 16\% & 18\% & 14\%  \\ \hline
\rowcolor[HTML]{FFFFFF} 
BNN   & 9\% & 11\% & 10\%  \\ \hline
\rowcolor[HTML]{FFFFFF} 
LR    & 17\% & 19\% & 26\% \\ \hline
\end{tabular}
\caption{\label{tab:error2} Averaged error relative to the width of the bounds from \Cref{tab:bounds} in \% of the parameter predictions plotted in \Cref{fig:performance} for NN, BNN and LR}
}
\end{table}

\begin{figure}[h!]
\begin{center}
\includegraphics[width=1\columnwidth]{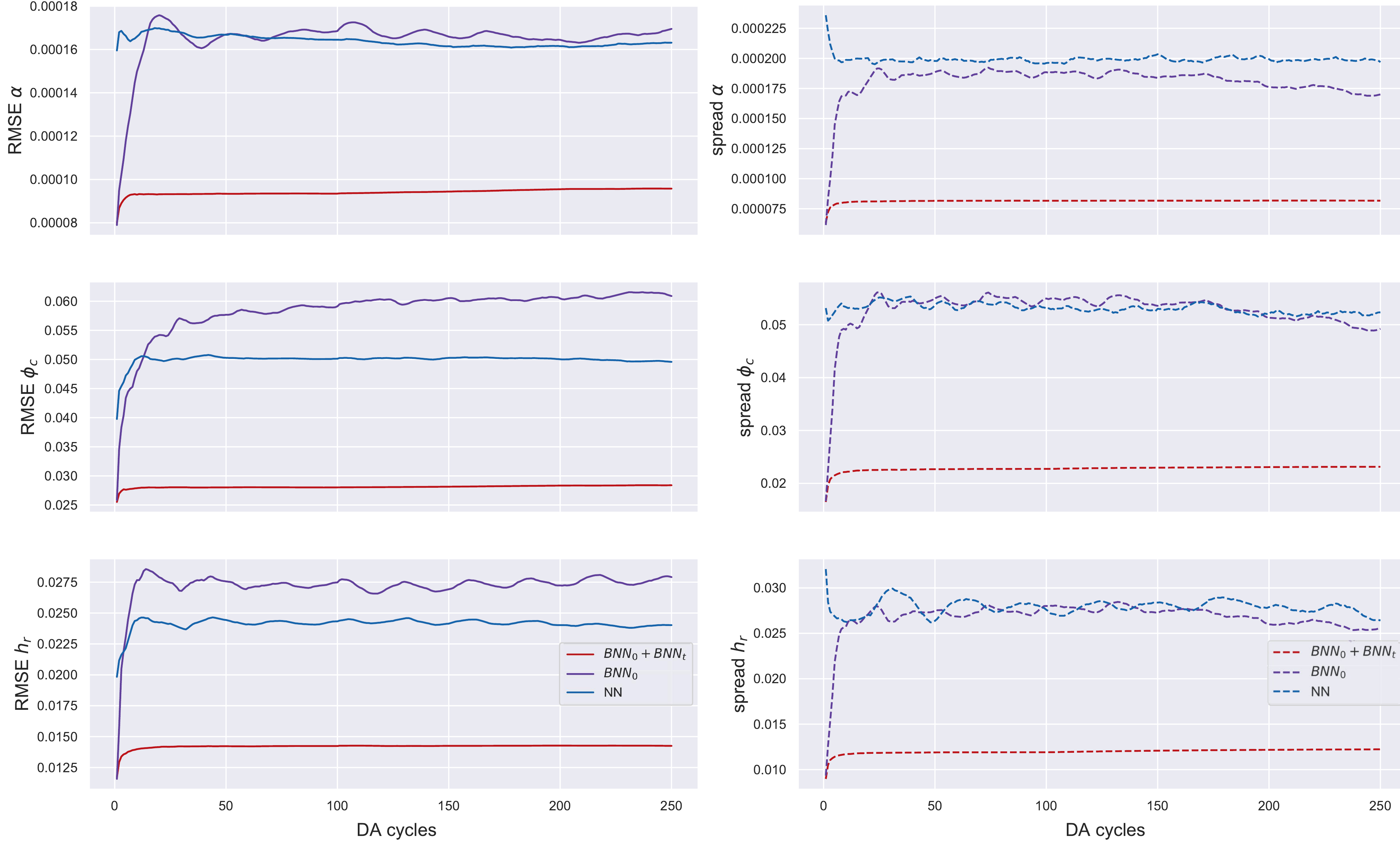}
\end{center}
\caption{Time evolution of RMSE (left) and ensemble spread (right) of parameter estimates against time in DA cycles with 50 analysis ensemble members averaged over 100 experiments}
    \label{fig:time_evolution}
\end{figure}

The remaining experiments presented in this subsection are conducted according to \Cref{chap:da_ml} and if not stated otherwise averaged over 100 individual experiments with different ground truth values. Since NN and BNN$_0$ are trained on snapshots of model states from one point in time, the question arises how they perform when using states from later points in time as the input and comparing them with BNN$_t$ which is constantly retrained over time. If the predictive power of the former do not significantly decrease it would only be necessary to train the ANNs once and they could then be used to predict parameters whenever necessary which would be computationally cheap. In \Cref{fig:time_evolution} the parameter RMSEs and parameter ensemble spreads of all ANNs are plotted against time in DA cycles where DA cycle = 0 corresponds to the time NN and BNN$_0$ were trained on and between two points on the x-axis lie 60 model time steps. The RMSE of the parameters is smallest at the beginning of the time evolution. After that, they grow for about 50 cycles and then oscillate around a relatively constant value  (\Cref{fig:time_evolution},left). Although BNN$_0$ outperforms the NN for the initial estimate at DA cycle = 0 over time the RMSEs increase a lot faster for BNN$_0$ compared to NN. For BNN$_0$+BNN$_t$ the RMSEs also increase over time, but to a lesser extent compared to the other methods. While the parameter spread of the BNNs increases which is in accordance to the increased RMSEs the spread of the NN actually decreases over time (\Cref{fig:time_evolution},right) which could be interpreted as the NN becoming more confident in its prediction over time.

\begin{figure}[h!]
\begin{center}
\includegraphics[width=1\columnwidth]{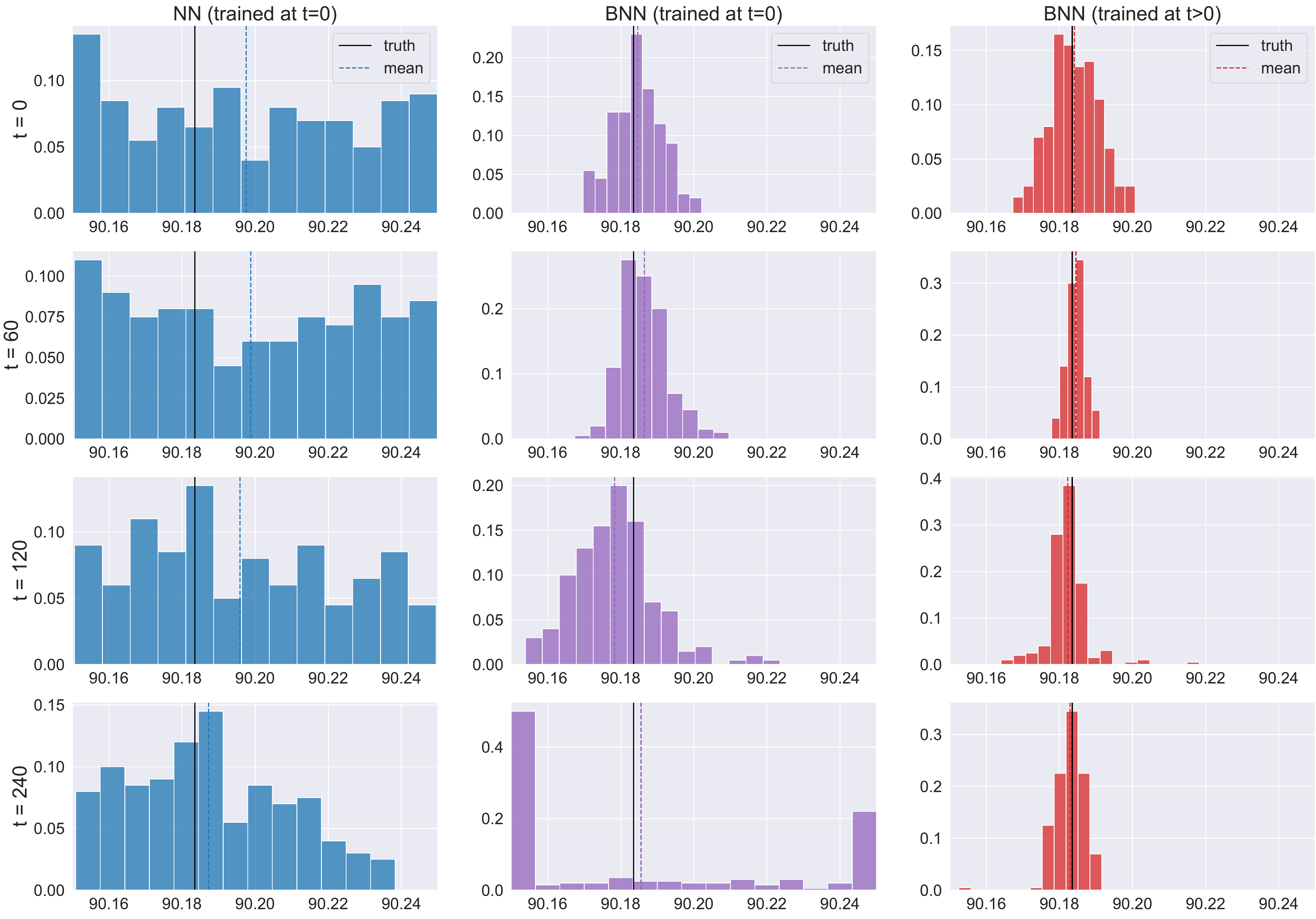}
\end{center}
\caption{Probability histograms of estimates for rain threshold \( h_r \) of one single experiment for NN (left), BNN$_0$ (middle), and BNN$_0$+BNN$_t$ (right) over time with 200 ensemble members}
    \label{fig:distribution_h}
\end{figure}

Histograms of the parameter estimation of \(h_r\) for one single experiment were plotted for the NN as well as for both BNN methods (\Cref{fig:distribution_h}). Since this is only a single experiment the results shown here are not statistically significant. However, they still illustrate the key differences between the methods. To also investigate the change of the distributions over time a histogram for each method was plotted from t=0 (DA cycle = 0) to t=240 (DA cycle = 4). While the NN estimates are spread out over the whole range with accumulations around the true value the estimates of the BNNs are close to Gaussian distributions with the mean near the true value and a small variance. Nonetheless, over time the prediction of BNN$_0$ spreads out and it loses its predictive power for \(DA cycle \geq 4\). Since BNN$_0$ starts predicting parameters that lie greatly outside of the bounds it was necessary to map the outliers to the bounds as otherwise the modified shallow water model could not run which explains the aggregations near the edges. Since this was not necessary for the other two methods the remaining experiments were only conducted using NN and BNN$_0$+BNN$_t$.

\begin{figure}[h!]
\begin{center}
\includegraphics[width=1\columnwidth]{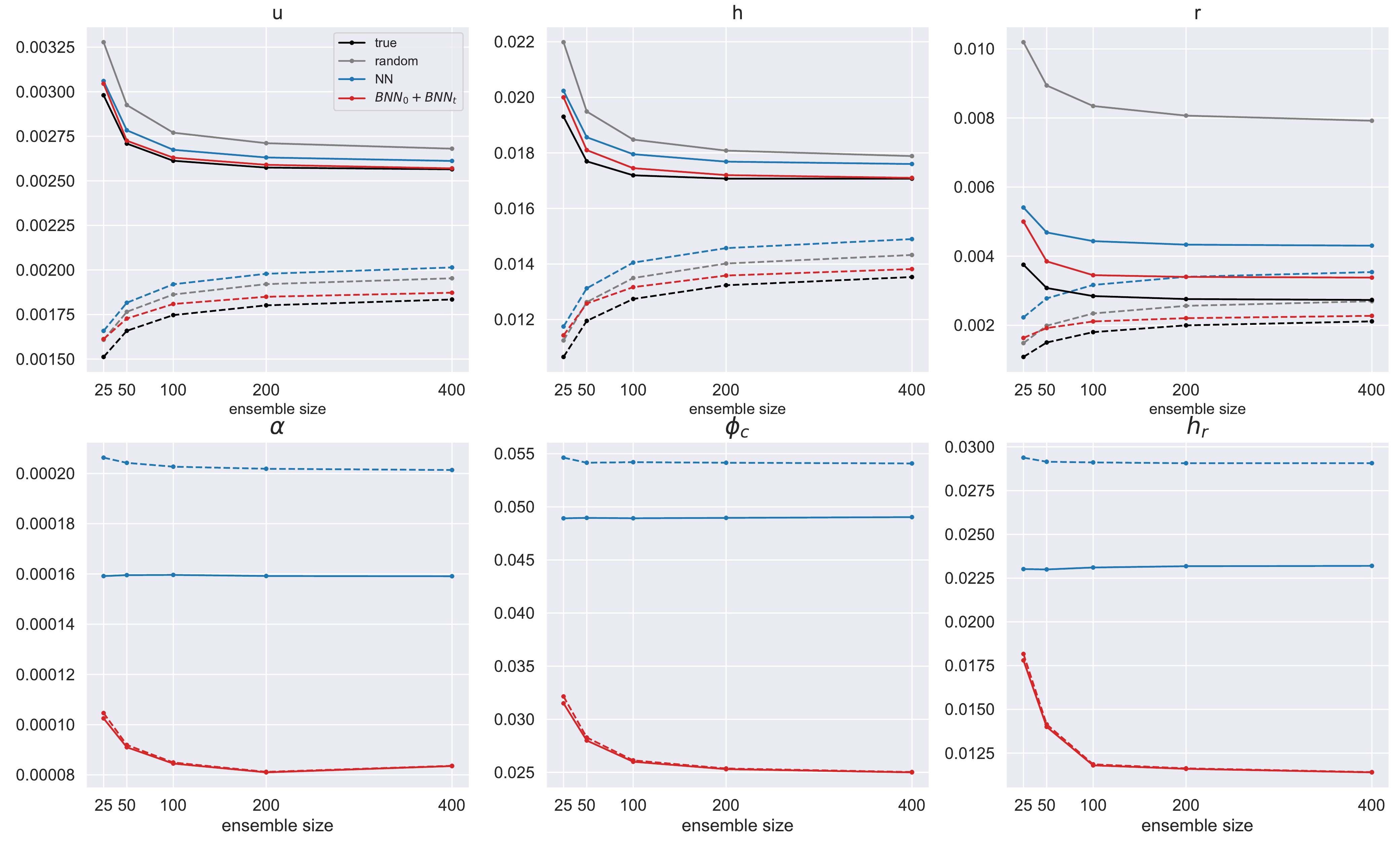}
\end{center}
\caption{RMSE (solid) and ensemble spread (dashed) of atmospheric variable estimates (upper panel) and parameter estimates (lower panel) against analysis ensemble size averaged over last 100 DA cycles of 100 experiments}
    \label{fig:ensemble_size}
\end{figure}

\begin{figure}[h!]
\begin{center}
\includegraphics[width=1\columnwidth]{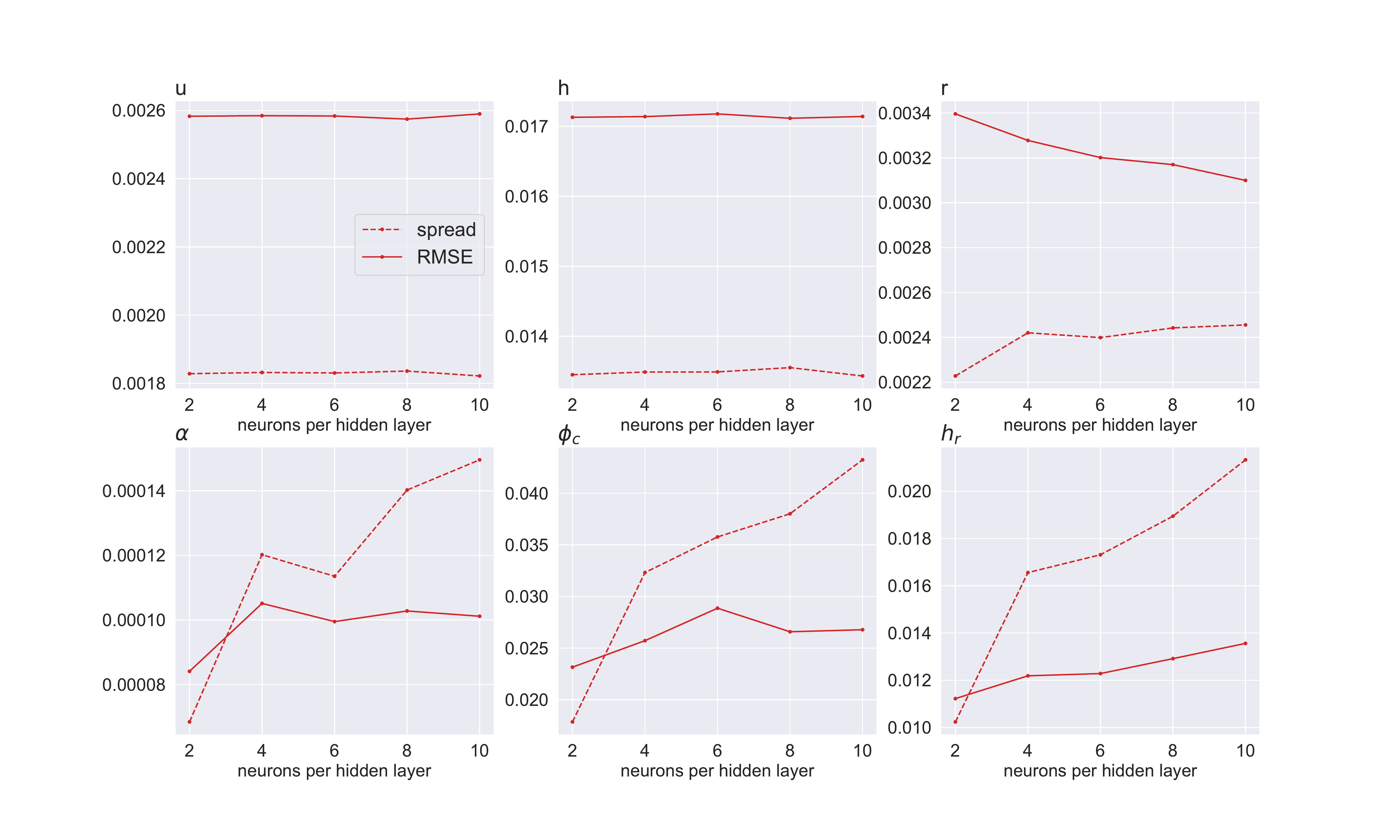}
\end{center}
\caption{RMSE (solid) and ensemble spread (dashed) of atmospheric variable estimates (upper panel) and parameter estimates (lower panel)  against network size in neurons per hidden layer averaged over last 100 DA cycles of 100 experiments using BNN$_0$+BNN$_t$ with 200 state ensemble members}
    \label{fig:network_size}
\end{figure}

\begin{figure}[h!]
\begin{center}
\includegraphics[width=1\columnwidth]{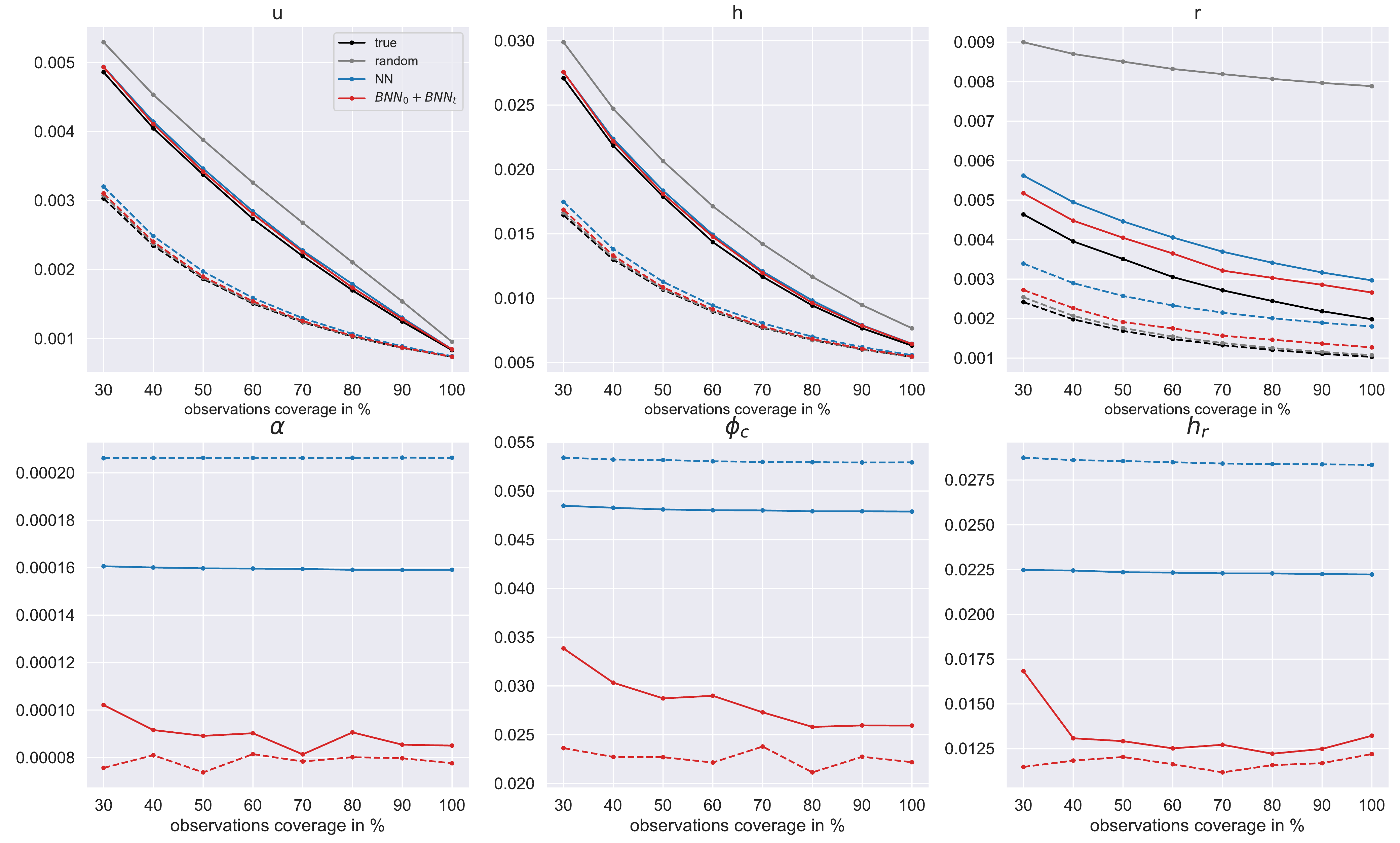}
\end{center}
\caption{RMSE (solid) and ensemble spread (dashed) of atmospheric variable estimates (upper panel) and parameter estimates (lower panel) against observation coverage with 50 ensemble members averaged over last 100 DA cycles averaged over 100 experiments}
    \label{fig:observation_coverage}
\end{figure}

\begin{figure}[h!]
\begin{center}
\includegraphics[width=0.7\columnwidth]{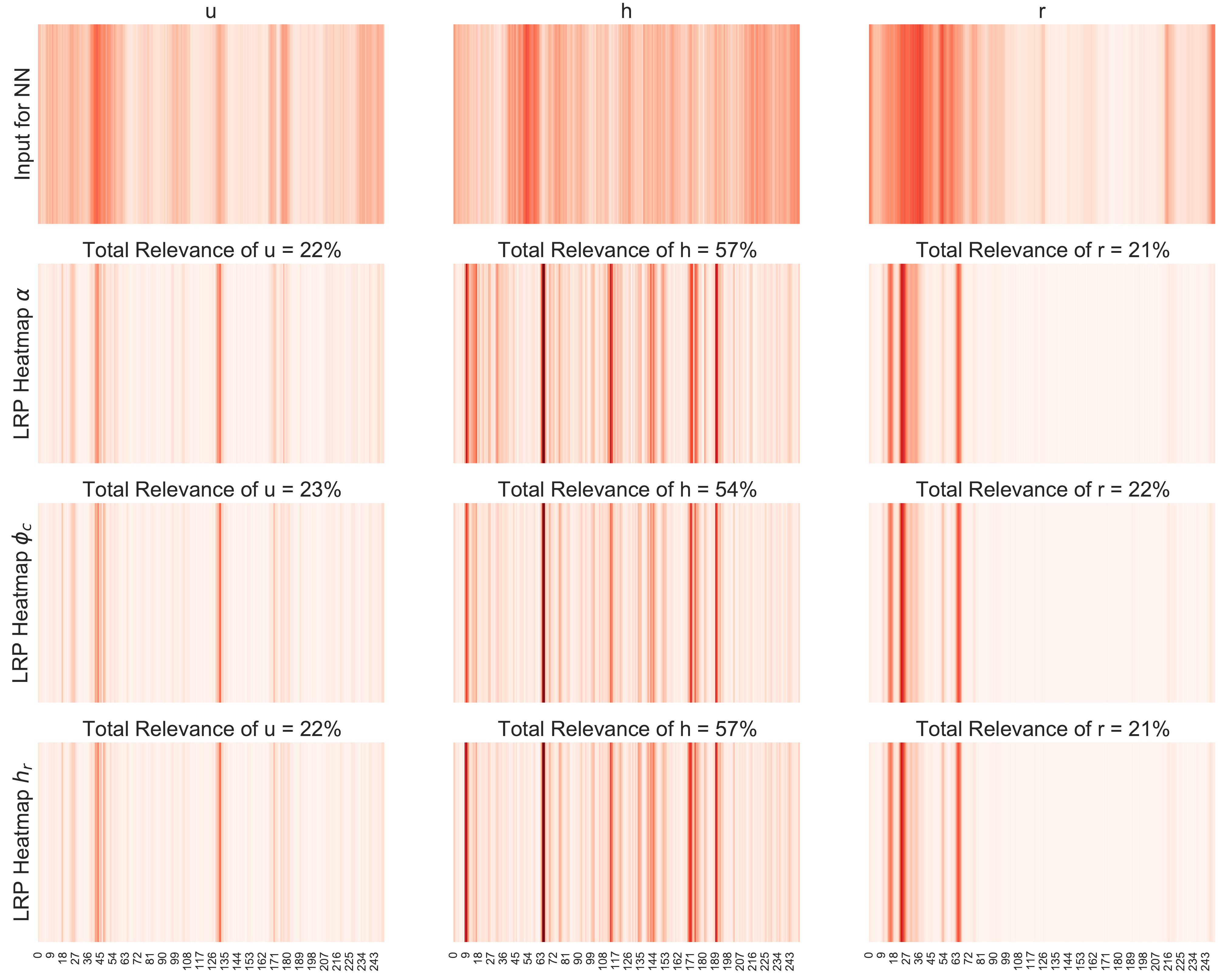}
\end{center}
\caption{Input (first row) and corresponding LRP heatmaps for $\alpha$ (second row), $\phi_c$ (third row) and $h_r$ (last row) against grid points of NN trained according to \Cref{chap:training} averaged over 100 experiments}
    \label{fig:lrp}
\end{figure}

\begin{figure}[h!]
\begin{center}
\includegraphics[width=0.7\columnwidth]{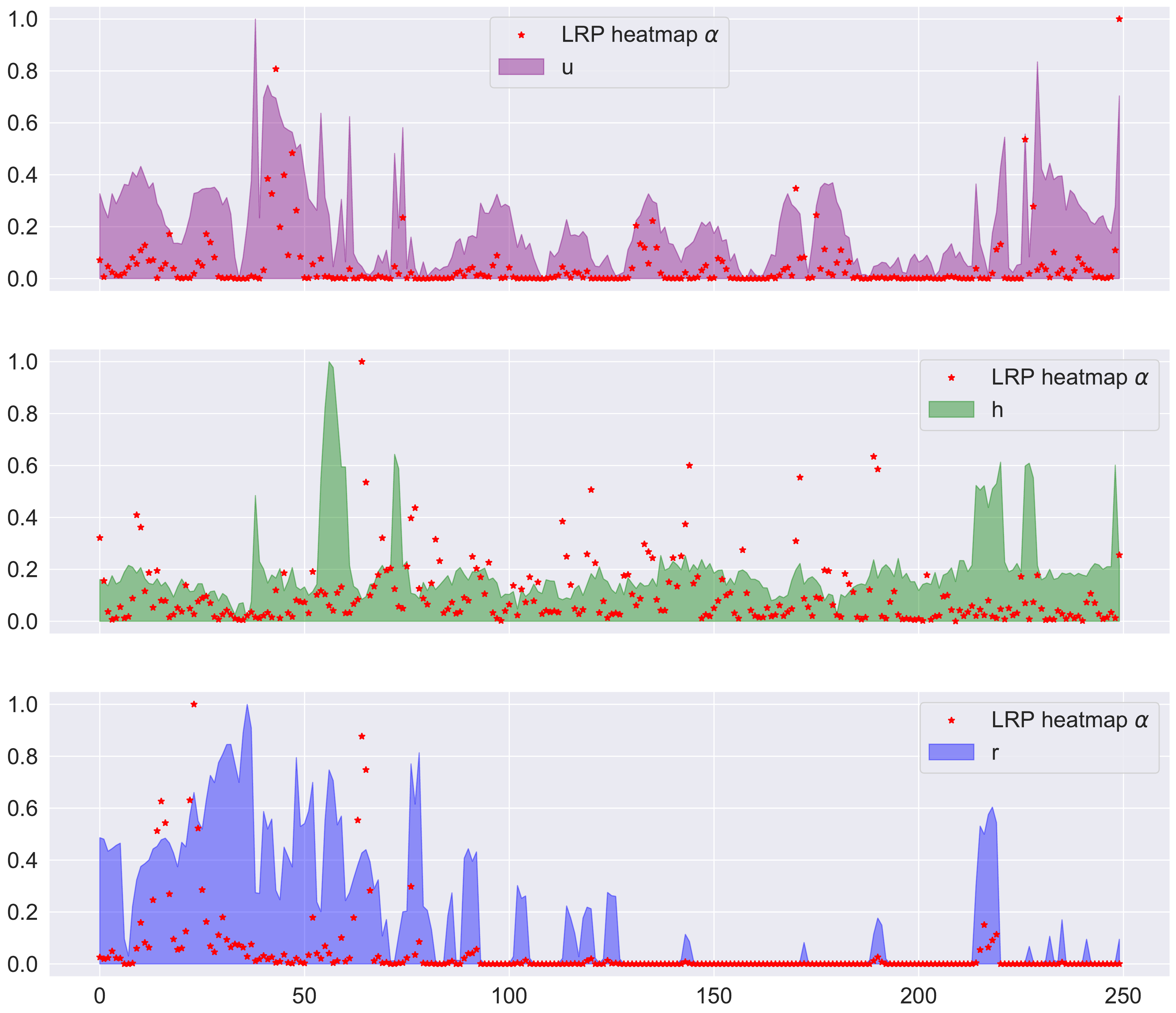}
\end{center}
\caption{Rescaled values of fluid velocity $u$ (upper panel), height $h$ (middle panel), rain $r$ (lower panel), and corresponding LRP heatmaps   $\alpha$ (red stars) of NN trained according to \Cref{chap:training} against all 250 grid points for a single experiment}
    \label{fig:lrp_alpha}
\end{figure}

\begin{figure}[h!]
\begin{center}
\includegraphics[width=0.7\columnwidth]{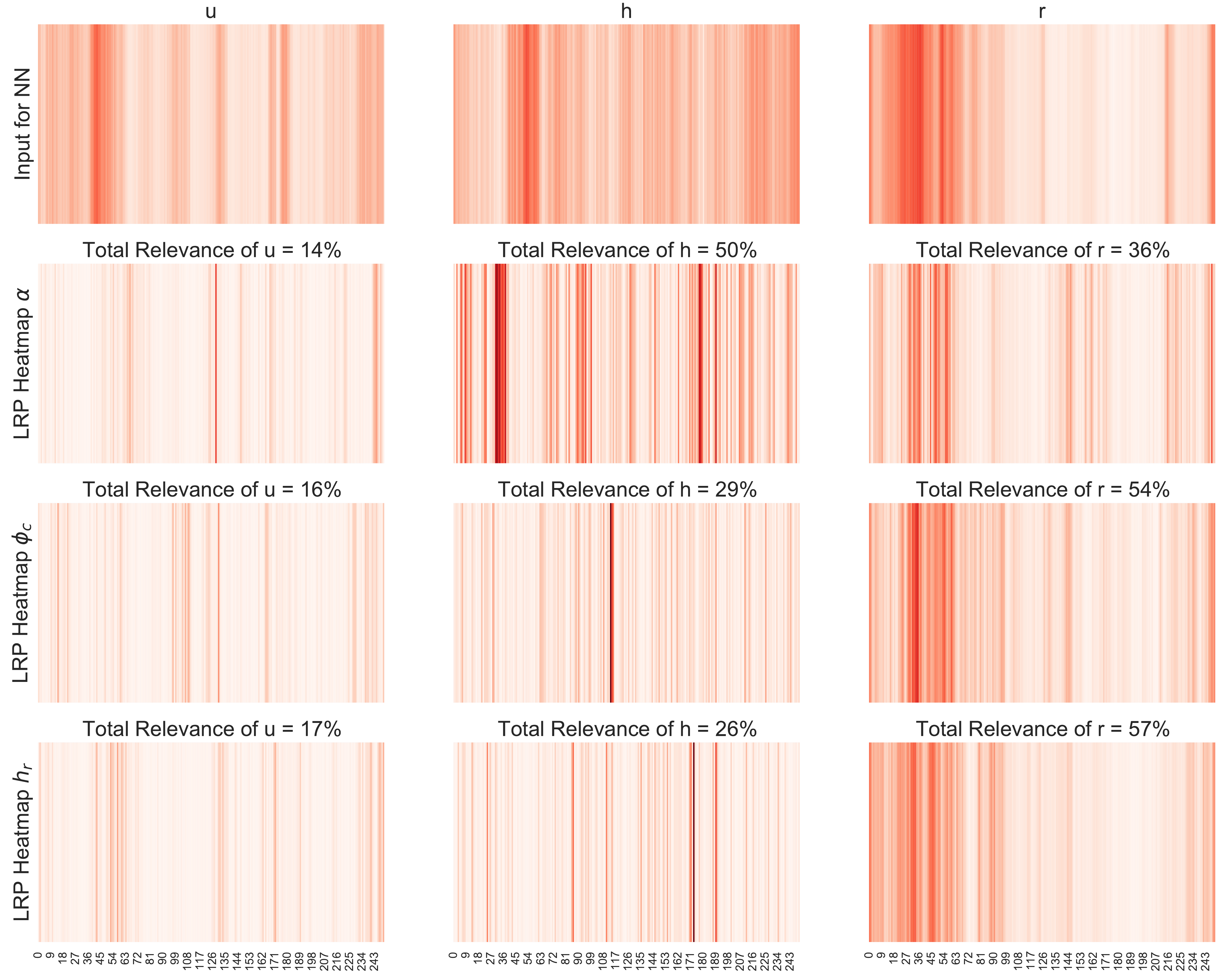}
\end{center}
\caption{Input (first row) and corresponding LRP heatmaps for $\alpha$ (second row), $\phi_c$ (third row) and $h_r$ (last row) against grid points of 3 individually trained NNs averaged over 100 experiments}
    \label{fig:lrp_individual}
\end{figure}

\begin{figure}[h!]
\begin{center}
\includegraphics[width=0.7\columnwidth]{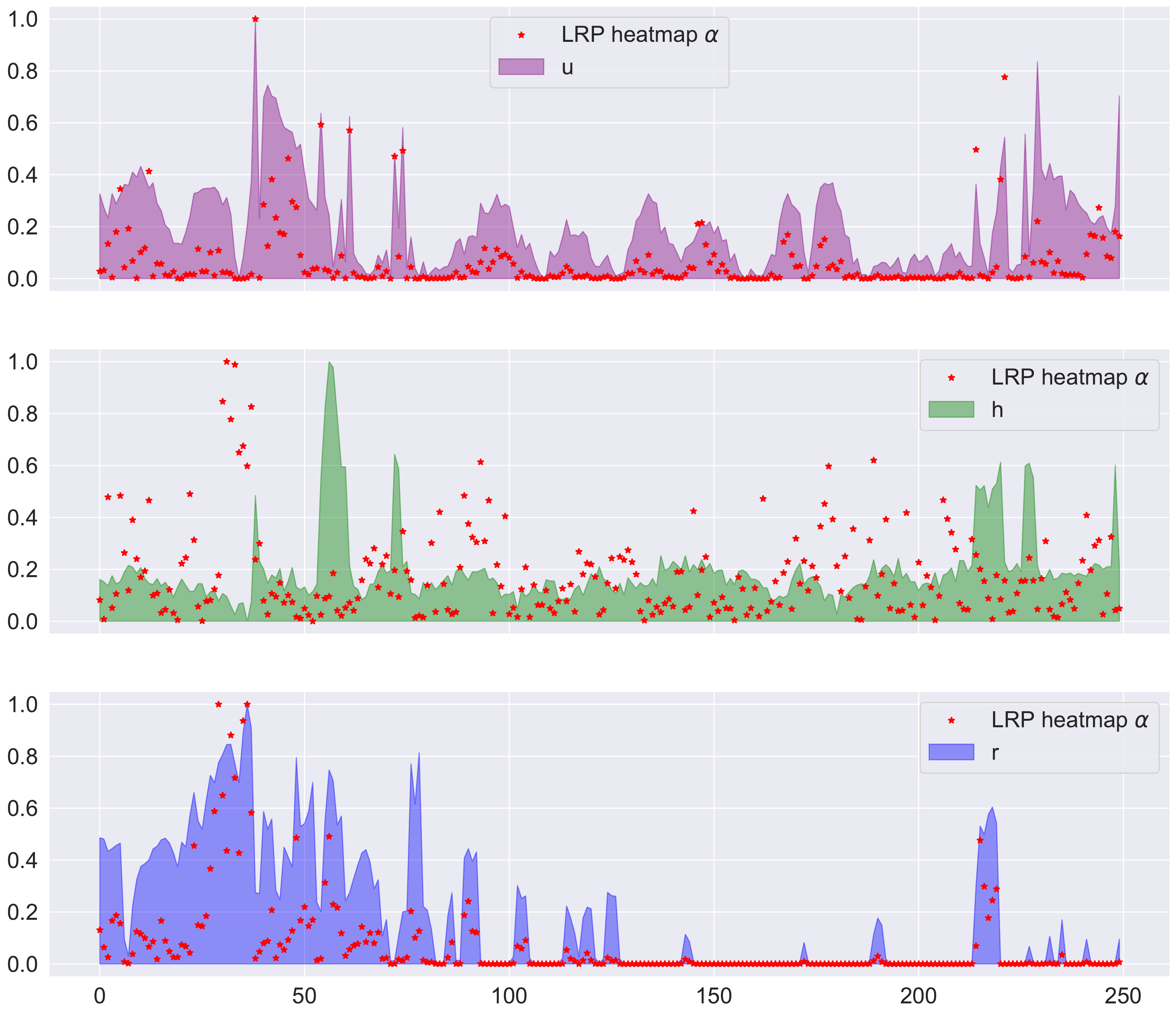}
\end{center}
\caption{Rescaled values of fluid velocity $u$ (upper panel), height $h$ (middle panel), rain $r$ (lower panel), and corresponding LRP heatmaps   $\alpha$ (red stars) of 3 individually trained NNs against all 250 grid points for a single experiment}
    \label{fig:lrp_alpha_individual}
\end{figure}

The sensitivity of the RMSE and ensemble spread of the atmospheric variable and the model parameter estimates to the number of state ensemble members (\Cref{fig:ensemble_size}) and observation coverage (\Cref{fig:observation_coverage}) is studied for NN and BNN$_0$+BNN$_t$ to compare their performance and statistics with the best (black) and worst (gray) case scenarios and to investigate the capabilities of the ANNs under sparse and noisy conditions. For all methods 100 experiments with 250 DA cycles each are conducted and averaged over the last 100 DA cycles. As expected, the RMSEs of the atmospheric variables decrease for all setups with an increase in state ensemble members (\Cref{fig:ensemble_size}) due to the samples being able to more accurately approximate the true Kalman filtering distribution. The BNN has lower RMSEs than the NN in all experiments and achieves results close to the best-case scenario for u and h for a large ensemble size of 400. However, the NN is more beneficial to the RMSE/spread ratio than the BNN which is likely due to the small parameter spread of the BNN. The parameter estimates of the NN do not exhibit any sensitivity to the state ensemble size. The sensitivity of the BNN can be explained due to the ensemble size directly controlling the training size and more training data usually increases the predictive capabilities of ANNs. For \(N_{ens}>100\) however, the RMSEs of the parameters seem to saturate which could be caused by the very small network size of only 2 neurons per hidden layer and might positively be influenced by increasing the neurons of the hidden layers. To test this hypothesis, network size sensitivity experiments are conducted in \Cref{fig:network_size}.
For the results shown in \Cref{fig:network_size} the same 100 experiments as before are conducted with the state ensemble size set to 200 and varied neurons per hidden layer. Contrary to our hypothesis, the parameter RMSEs increase with an increase in neurons as does the spread. This surprisingly has a positive effect on the rain r where the RMSE decreases while its spread increases. The velocity u and fluid height h on the other hand show no sensitivity to the network size. \Cref{fig:network_size} indicates that not only the accuracy of the parameter estimation but also its spread is relevant for the state error.
The NN and BNN$_0$ are trained on the full grid, but observations in real-life settings are usually sparse. Therefore, their sensitivity to the \emph{observation size} was investigated, which is here defined as the percentage of observed grid points. Instead of observing only those grid points whose rain values exceed a certain threshold, as in the previous experiments, percentages corresponding to the values of the x-axis in \Cref{fig:observation_coverage} of random grid points were observed. The RMSEs of the atmospheric variables, especially those of u and h, show a strong sensitivity on the observation size and decrease with more observations available. The parameter RMSEs of the NN, on the other hand, show almost no sensitivity at all. The BNN exhibits a slightly stronger sensitivity compared to the NN up until around 60\% of available observations. This low sensitivity on the observation size indicates that although the ANNs are trained on the whole grid, only a small number of grid points could actually be relevant for the ANNs' prediction. To investigate this hypothesis the LRP algorithm described in \Cref{chap:diagnostics} is utilized in \Cref{fig:lrp,fig:lrp_individual,fig:lrp_alpha,fig:lrp_alpha_individual} with the NN since to the best of our knowledge, LRP has not been applied to BNNs so far.
For \Cref{fig:lrp} LRP was applied to all three parameters \(\alpha, \phi_c\) and \(h_r\) for 100 inputs and averaged. The inputs used in these experiments are observations taken from the true atmospheric state of fully observed grids. To better compare the NN inputs (observations of the state) with their corresponding LRP heatmaps, the values of u,h,r and the LRP heatmap values were rescaled between 0 and 1 and plotted together as heatmaps. The x-axis represents the 250 grid points while the y-axis has no meaning and just provides a spatial dimension so that the colors of the heatmap are visualized better. Darker red tones correspond to higher values and thus represent grid points that were more relevant for the NN's prediction. The total relevances plotted in \Cref{fig:lrp} are simply the LRP heatmap values of one parameter for a certain atmospheric variable summed and divided by the total sum of LRP heatmap values for that parameter. Even though all experiments conducted so far indicate that r is the most sensitive variable to the parameter estimation, for the NN h is the most relevant variable while u and r exhibit roughly the same relevance. It is also surprising that even though the results plotted in \Cref{fig:lrp} are averaged over many experiments, one can still determine distinct lines over certain grid points. These distinct lines indicate that the NN uses only a small number of grid points to make its prediction instead of the whole grid. Using only a small number of important grid points as the input would in turn decrease the number of learnable parameters and might result in the need for much smaller training sizes. This finding would also explain why the sensitivity on the observation size is so low. The heatmaps of u and r look very similar to the input indicating that those grid points with strong winds and rain are especially relevant for the NN. The heatmaps of h however look very different from their input. In \Cref{fig:lrp_alpha} the heatmaps of a single experiment are plotted for all three parameters to investigate which grid points are relevant and if the heatmaps of the three parameters indeed look as similar as \Cref{fig:lrp} indicates.
Instead of plotting the LRP heatmaps and the atmospheric variable values as heatmaps, in \Cref{fig:lrp_alpha} they are visualized as graphs with the x-axis corresponding to the grid points and the y-axis corresponding to the rescaled values. The shaded areas represent the values of u,h,r while the red stars are the heatmap values for \(\alpha\) of the corresponding inputs. Visualizing the LRP outputs this way makes it even more obvious that grid points with strong winds and rain are very relevant for the NN, although there does not seem to be a distinct relationship between h and its corresponding LRP heatmap. If one plots the LRP heatmap \( \alpha \) of h together with the rain (not shown), it seems like the relevant grid points of h are the ones where it is in fact raining. This would explain why simply interpolating the observations and using these to estimate the parameters, as was done for in  \Cref{fig:time_evolution}, works so well. From a physical standpoint, it is rather surprising that the heatmaps in \Cref{fig:lrp} look so similar for all three parameters. To check if this is simply due to the fact that they are predicted simultaneously, the same experiments as in \Cref{fig:lrp} are repeated with three individually trained NNs, one for each parameter. For the results in \Cref{fig:lrp_individual} three training and test data-sets are generated: each time keeping two of the parameters constant while varying the parameter that is wished to be estimated. In this setup, there is a clear distinction between the heatmaps of the three individual parameters and also between the heatmaps of \Cref{fig:lrp} and \Cref{fig:lrp_individual}. The relevances shift from the fluid velocity u and height h towards the rain r for all three parameters, especially for \(\phi_c\) and \(h_r\) where the rain is now the most relevant variable. For the rain removal rate \(\alpha\) the most relevant variable is still h. While in \Cref{fig:lrp} the relevances are concentrated on a few single grid points they are more spread out over the grid in \Cref{fig:lrp_individual}. To investigate if this spread is due to the averaging over many experiments the same plot as in \Cref{fig:lrp_alpha} is created for the three individually trained NNs for the same  experiment. When comparing \Cref{fig:lrp_alpha} with \Cref{fig:lrp_alpha_individual} it becomes apparent that when the NNs are trained for each parameter individually, almost all rainy grid points are now relevant for the NN's prediction instead of just a select few.

\section{Conclusion}\label{sec:con}
In this study two types of ANNs are trained to estimate the tunable model parameters of the convective-scale modified shallow water model \citep{wuersch14}. In the perfect model experiments the NN as well as the BNN are able to decrease the state errors of the atmospheric variables compared to the case were no parameter estimation is applied. The largest reduction of the state error is ultimately found in the rain \(r\). Furthermore, the ANNs investigated here provide tools to quantify the uncertainty of the parameter estimation which increases the ensemble spread of the state analysis and forecast while decreasing their RMSEs. Interestingly, even though the rain exhibits the largest sensitivity on the parameter estimation, the LRP algorithm shows that the fluid height h was the most relevant variable for the NN's prediction. In summary, the BNN produced more accurate estimates while needing less training time and hyperparameter tuning.\\
All experiments conducted in this study assume parameters that are constant in time and space. Future work testing the ANNs' ability to estimate local and temporal parameters is therefore required. Furthermore, the training data utilized in this study are snapshots of the grid at one point in time. Alternatively, one could investigate the use of different features such as time series of one or more grid points. By utilizing not only atmospheric variables as input features but also 'climatological predicators', such as latitude, longitude, time of the day and  month, \citep{bonavita20} show that the the predictive abilities of the ANNs are greatly enhanced. Providing the ANN with information on the geographical location, diurnal cycle, and seasonal cycle during the training is an interesting approach that could have potential benefits for the parameter estimation problem as well. It should be noted that the training data, as well as the observations, are generated by the same simplistic model and it is not clear how well the ANNs' predictive ability translates to more complex models and real observations. Before testing them in more realistic scenarios, it is necessary to scale down their demand for large training sizes, possibly by reducing the number of input features or number of hidden layers as demonstrated here. Indeed, the LRP heatmaps show that if all parameters are estimated simultaneously the NN makes use of only a few select grid points.\\
Another possibility to address the here mentioned challenges would be to investigate an alternative kind of stochastic ANN. \citet{leinonen20} succesfully train a stochastic generative adversarial network (GAN) to downscale time-evolving images of atmospheric fields from low to high resolution. The GAN trained in \citet{leinonen20} consists of convolutional and recurrent layers and is able to predict images larger than those it was trained on and can predict longer time series than the sequences used for training. This reduces the need for large training sizes and offers the possibility for offline training.\\ 
In the studied test case, with perfect model assumptions and enough training data, the ANNs were able to estimate the unknown model parameters and quantify their uncertainty more accurately than a simple linear regression, even under sparse and noisy conditions. Including the parameter estimates obtained from the ANNs in the DA cycle resulted in reduced state errors and increased ensemble spreads compared to the case without parameter estimation and unknown parameters.

\section{Acknowledgements}\label{acknowledgements}
The research leading to these results has been done within the subproject B6 of the Transregional Collaborative Research Center SFB / TRR 165 “Waves to Weather” funded  by  the  German Research  Foundation  (DFG). T. Janji\'{c} is  also grateful to the DFG for funding of her Heisenberg Award (DFG JA1077/4-1).

\section*{Conflict of interest}
The authors declare that they have no conflict of interest.

\selectlanguage{english}
\bibliographystyle{unsrtnat}
\bibliography{references}

\end{document}